# Decay of Relevance in Exponentially Growing Networks

Jun Sun
Institute WeST, Universität
Koblenz–Landau
Koblenz, Germany
junsun@uni-koblenz.de

Steffen Staab*
Institute WeST, Universität
Koblenz–Landau
Koblenz, Germany
staab@uni-koblenz.de

Fariba Karimi
GESIS — Leibniz-Institut für
Sozialwissenschaften
Cologne, Germany
fariba.karimi@gesis.org

## ABSTRACT

We propose a new preferential attachment–based network growth model in order to explain two properties of growing networks: (1) the power-law growth of node degrees and (2) the decay of node relevance. In preferential attachment models, the ability of a node to acquire links is affected by its degree, its fitness, as well as its *relevance* which typically decays over time. After a review of existing models, we argue that they cannot explain the above-mentioned two properties (1) and (2) at the same time. We have found that apart from being empirically observed in many systems, the exponential growth of the network size over time is the key to sustain the power-law growth of node degrees when node relevance decays. We therefore make a clear distinction between the *event time* and the *physical time* in our model, and show that under the assumption that the relevance of a node decays with its age $\tau$, there exists an analytical solution of the decay function $f_R$ with the form $f_R(\tau) = \tau^{-1}$. Other properties of real networks such as power-law alike degree distributions can still be preserved, as supported by our experiments. This makes our model useful in explaining and analysing many real systems such as citation networks.

## CCS CONCEPTS

• **Networks** → *Network architectures*; • **Human-centered computing** → Social network analysis;

## KEYWORDS

network growth model, preferential attachment, decay of relevance

## 1 INTRODUCTION

Network growth models try to explain the evolution of numerous types of networks, such as social networks, citation networks and the World Wide Web. These networks — despite of their heterogeneity in terms of their purposes, their origins and the natures of their nodes and links — tend to have certain phenomena in common. At the macro level, most real networks appear to be a scale-free structure indicated by the presence of power-law like node degree distributions [6, 13]. Many networks grow exponentially, especially as long as each node is able to attract yet another, similar sized group of nodes as the ones before it [7]. Such growth might eventually slow down due to the limited number of potential nodes [15], for example when a social network already covers most of the population. At the micro level, nodes in the networks have varying abilities to attract new nodes, thus having different degree growth curves. Besides, new nodes "prefer" to "attach" to existing nodes with higher degrees [1]. A node's ability to attract new nodes is also affected by its intrinsic *fitness* [3] and its *relevance* which typically decays over time [12].

Preferential attachment models including the Barabási-Albert model [1] or the Bianconi-Barabási model [3] have been proposed and have successfully enlighted one important connection between the two levels of phenomena: The *rich-get-richer* effect resulting from preferential attachment at the micro level leads to the scale-free nature of the network at the macro level. Other preferential attachment models have been proposed to deal with individual aspects of phenomena such as the accelerated growth of the network [7] and the decay of node relevance [11, 12].

In our study, we try to reveal another connection among the growth of network sizes, the growth of node degrees and the decay of node relevance. More specifically, we argue that existing preferential attachment models cannot well explain the micro level phenomenon: power-law growth of node degrees under the decay of node relevance. We therefore propose a new preferential attachment–based network growth model that shows the exponential growth of the network size at the macro level — apart from being empirically observed — is the key to sustain the power-law growth of node degrees when node relevance decays. We show analytically that the node's age $\tau$ and the node relevance decay function $f_R(\tau) = \tau^{-1}$ may connect the micro and macro levels of phenomena. To make our model work with a wider range of networks, we also discuss situations where the exponential growth slows down.

First, we review basic notations of networks and some existing network growth models in Section 2. We then look into real world examples of networks in Section 3, and show various empirical observations which existing models cannot explain well at the same time. In Section 4, we propose our model on the basis of these observations, and analytically show its self-consistency. In order to evaluate our model, in Section 5 we generate synthetic networks using our model with different parameters, and compare the generated networks with real world networks. In Section 6, we conclude with some discussions and point to potential future work.

---
*Also with Institute WAIS, University of Southampton.

Table 1: List of Notations

| Sym. | Meaning |
|---|---|
| $s$ | the event time, whose advance is driven by events in the network. When written as $s_i$ with a subscript $i$, it represents the event time that node $i$ joins the network; |
| $t$ | the physical time. $t_i$ represents the physical time that node $i$ joins the network; |
| $\tau$ | the age of a node in physical time. For mathematical simplicity we let $\tau$ start at 1 when the node joins the network. $\tau_i$ represents the age of node $i$; |
| $\eta$ | the fitness of a node that remains constant over time (see Section 2.3). $\eta_i$ represents the fitness of node $i$; |
| $\rho(\eta)$ | the probability density function characterizing the fitness distribution of nodes in the network; |
| $\Pi$ | the preferential attachment probability of a node (see Section 2.2). $\Pi_i$ represents the preferential attachment probability of a new node to attach to node $i$; |
| $T$ | the age of the whole network in physical time; |
| $k$ | the (in)degree[1] of a node. When written as $k_i$ with a subscript $i$, it represents the degree of node $i$. $k_i$ is a function of $s$, $s_i$ and $\eta_i$: $k_i(s, s_i, \eta_i)$, or a function of $\tau_i$ and $\eta_i$: $k_i(\tau_i, \eta_i)$; |
| $f_R(\tau)$ | the relevance decay function which is a monotonically decreasing function with regard to the node age $\tau$ that characterises the decay of node relevance (see Section 2.4); |
| $p(k)$ | the probability density function characterising the degree distribution of the network. |

## 2 RELATED WORK

In this section, we review some existing preferential attachment models, namely the Barabási-Albert model [1], the Bianconi-Barabási model [3] and the relevance model [12].

### 2.1 Notations in Preferential Attachment Models

In order to compare preferential attachment models with each other and with our model, we have come up with a common notation displayed in Table 1 that we will use consistently in the remainder of this paper.

Note that conventionally $t$ has been used to represent timesteps in the three existing models that we discuss in this section, and indeed there is no need to distinguish $s$ and $t$ if a uniform growth of the network size is assumed. In this paper we lift this assumption and explicitly distinguish $s$ and $t$.

---

[1] Since in our model the outdegree of a node is trivially constant, without explicitly mentioning, the term *degree* always stands for indegree.

### 2.2 Barabási-Albert Model

Albert and Barabási have proposed the well known Barabási-Albert model for network growth [1]. The model starts with one node with a self loops at timestep $s = 1$. At each later timestep $s > 1$, a new node joins the network and creates a constant number of $m$ links to existing nodes, each with the *preferential attachment probability* $\Pi_i$ proportional to the degree $k_i$ of the existing node $i$ [5]:

$$\Pi_i = \frac{k_i}{\sum_j k_j}. \quad (1)$$

When the network is large enough to apply the continuum approximation, the model analytically leads to two properties of the network.

(1) At the micro level, the power-law growth of node degrees:

$$k_i(s, s_i) \sim \left(\frac{s}{s_i}\right)^{1/2}, \quad (2)$$

where $s_i$ is the timestep when node $i$ joins the network.

(2) At the macro level, the power-law degree distribution:

$$p(k) \sim k^{-3}. \quad (3)$$

### 2.3 Bianconi-Barabási Model

The Barabási-Albert model successfully explains the scale-free property of networks. However, it predicts that the degrees of all nodes grow with a power function with the same exponent $\frac{1}{2}$, therefore old nodes tend to remain more popular than late comers. However in reality degrees of nodes grow with different rates, and often we see new nodes get more popular than old ones. Bianconi and Barabási have proposed to use the *fitness* value $\eta$ to quantify the ability of a node to acquire new links in the network [3]. When a node joins the network, it is assigned a fitness value $\eta$ which does not change over time. Being an extension of the Barabási-Albert model, the fitness model modifies the preferential attachment probability $\Pi_i$ that a newly arriving node links to an existing node $i$ to be proportional to the product of the degree $k_i$ and fitness $\eta_i$ of node $i$:

$$\Pi_i = \frac{\eta_i k_i}{\sum_j \eta_j k_j}. \quad (4)$$

The degree growth of a node still follows a power-law:

$$k_i(s, s_i, \eta_i) \sim \left(\frac{s}{s_i}\right)^{\beta(\eta_i)}, \quad (5)$$

but with an exponent $\beta(\eta_i)$ that is proportional to its fitness. The degree distribution of the network is however determined by the fitness distribution $\rho(\eta)$. For most fitness distributions, the resulting degree distribution still reflects the scale-free property of real world networks, although not being a perfect power-law.

### 2.4 Relevance Model

The Bianconi-Barabási model does not consider the decay of interest or *relevance* of nodes (e.g., scientific papers) that is often observed in reality. In the relevance model [12], the temporal decay of node relevance is modelled by a monotonically decreasing function $f_R(\tau)$ where $\tau$ is the age of a node (here in event time). The preferential

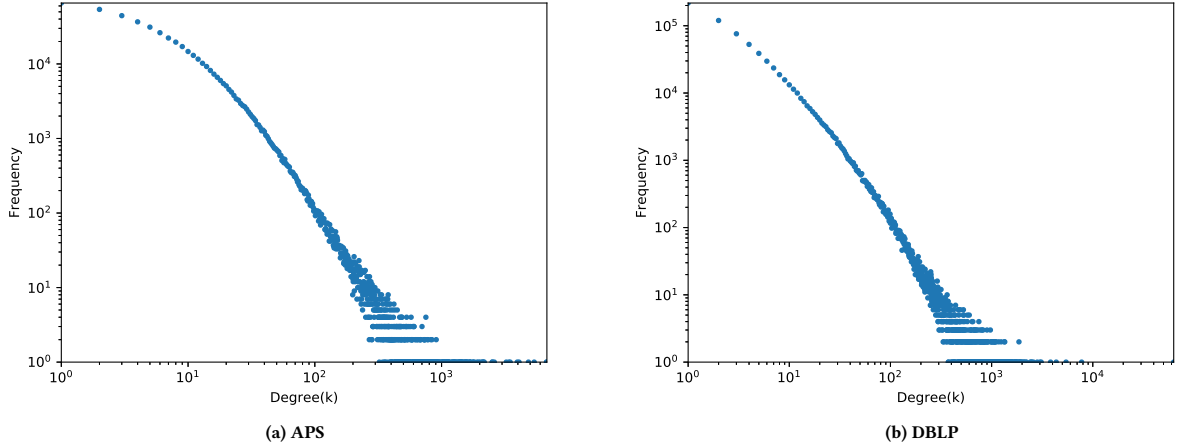

(a) APS					(b) DBLP

Figure 1: The degree distributions of the (a) APS and (b) DBLP citation networks. Estimated power-law exponents (for $k > 5$): 1.906 for APS and 1.947 for DBLP.

attachment probability is thus:

$$\Pi_i = \frac{\eta_i k_i f_R(\tau_i)}{\sum_j \eta_j k_j f_R(\tau_j)}. \qquad (6)$$

Medo et al. have reported that the relevance model can result in realistic degree distributions, such as exponential, log-normal and power-law distributions, depending on the input parameters [12].

## 3 EMPIRICAL OBSERVATIONS

In this section, we first introduce the two datasets that we use report the empirical findings on them. We then argue that existing models cannot well explain some observations.

### 3.1 Datasets

We use two datasets in this paper, each consisting of a citation network of research papers in two different diciplines.

The APS dataset is provided by the American Physical Society (APS)[2]. It contains a citation network consisting of 564,517 papers published in APS journals from 1893 to 2015, and 6,715,562 citations to other papers within the network.

The DBLP dataset is extracted by Tang et al. in the ArnetMiner project [16]. It contains 3,272,991 papers which were published from 1936 to 2016 and indexed in the DBLP Computer Science Bibliography [9], and 8,466,859 citations to other papers within the network. Since the original DBLP index is incomplete, there exist missing links in DBLP.

Both datasets contain the publication dates of papers. APS has a monthly granularity of dates, and DBLP has a yearly granularity.

[2] https://journals.aps.org/datasets

### 3.2 Scale-free Degree Distribution

Figure 1 shows the degree distribution of the two networks. Several observations can be made from the plots. The two distributions display power-law like behaviour, although not for very small degrees (under about 10) in APS. Additionally, the curve of APS has a well-defined long tail. The curve of DBLP displays an exponential cut-off when the degree is large (above around 200), which is also observed in many other networks [4]. We use the method by Clauset et al. [6] to estimate the power-law exponents of the two distributions and have found that they have comparable values (1.906 for APS and 1.947 for DBLP).

### 3.3 Exponential Growth of the Network Size

The growth of real systems is rarely linear. In the two datasets that we use, the network size $s$ (defined as the total number of publications) exhibits exponential growth with regard to the time $t$ (see Figure 2), as is observed in many systems [8] especially in their early stages. This exponential relationship $s = e^{\alpha t}$ can be explained by the Malthusian growth model, where $\alpha$ is the growth rate.

Being non-linear to the *physical time* $t$, the network size $s$ can be seen as the *event time* whose addition is driven by the arrival of new nodes. The non-linear relationship between $s$ and $t$ is however not considered in the models discussed in Section 2. In this paper we make a clear distinction between them.

### 3.4 Power-law Growth of Node Degrees

In the Bianconi-Barabási model, the growth of a node degree follows a power-law, with an exponent monotonic to the node's fitness.

We also observe power-law growths of node degrees in our datasets. Since the growth of individual nodes can be highly influenced by randomness, instead, we first group the papers in our datasets by their final citations received[3], and then plot the average growth of their degree in Figure 3. We observe linear curves in the log-log plot, but with different slopes. This indicates the power-law growth of node degrees with a fitness dependent exponent, as in the Bianconi-Barabási model.

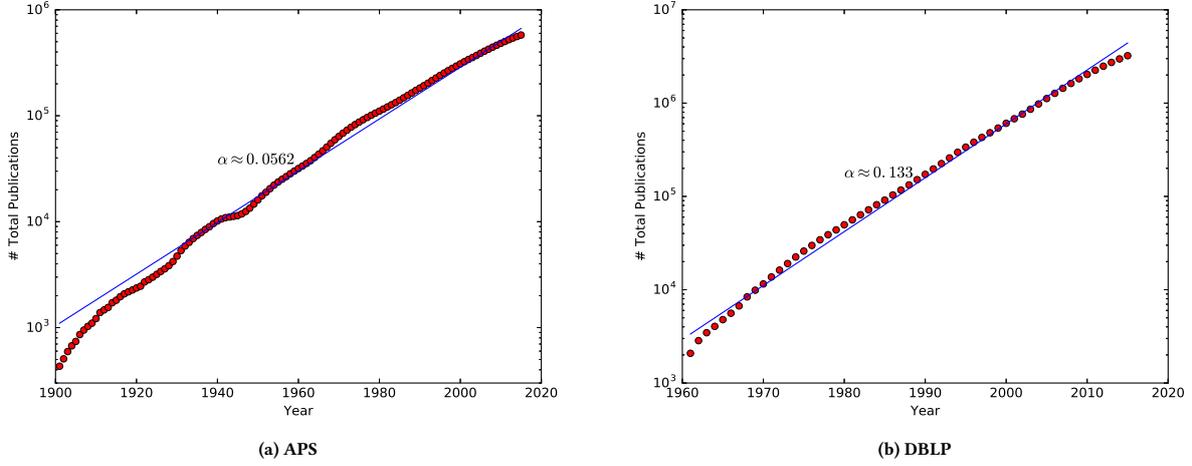

Figure 2: Total number of publications till each year in (a) APS and (b) DBLP. In each plot, the X-axis shows the physical timeline $t$ in years; while the Y-axis shows the total number of publications $s$ until the end of the corrsponding year in a logarithmic scale. The two approximately straight lines suggest the exponential growth of network sizes, i.e. $s \simeq e^{\alpha t}$.

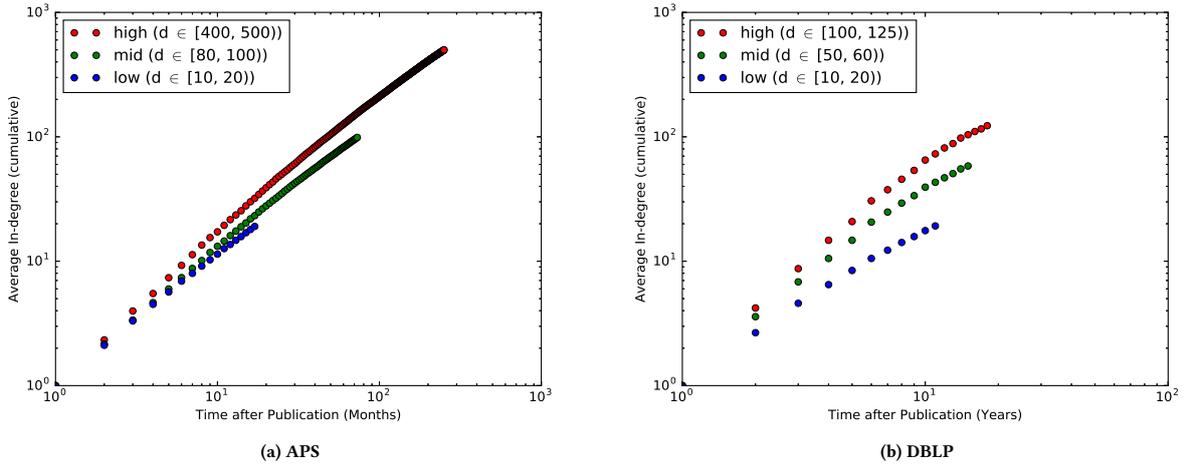

Figure 3: The average growth curve of number of citations of papers in (a) APS and (b) DBLP, grouped by their final indegrees.

### 3.5 Problems in Existing Models

In the Bianconi-Barabási model, the degree $k_i$ of a node $i$ with fitness $\eta_i$ at timestep $s$ can be written as Equation 5: $k_i \sim (s/s_i)^{\beta(\eta_i)}$. This indicates that the degree of a newer node $u$ will grow slower than an older one $v$, even if they have the same fitness, because $s/s_u < s/s_v$. However when we observe the degree growths of nodes grouped by their publication time in Figure 4, we see similar average growth curves of papers published in different time periods, which indicates that the growth depends on the age of the node $\tau_i$ but not $s/s_i$ as in the Bianconi-Barabási model, i.e. we have empirically:

$$k_i(\tau_i, \eta_i) \sim \tau_i^{\beta(\eta_i)}. \quad (7)$$

In Section 5.2 we discuss the difference between the two kinds of power-law growths (Equation 5 and 7) in detail.

Moreover, the initial derivation resulting in Equation 5 is based on linear network growth and no decay of relevance. If we solely consider (1) the decay of node relevance as in the relevance model, or solely consider (2) the exponential growth of the network size, the power-law growth of node degrees will be lost. In our model we propose to combine the two.

---
[3]Finding the actual fitness values is tricky. Mariani et al. have shown that the indegree is a good measurement to rank nodes by their fitness in such temporal networks [10].

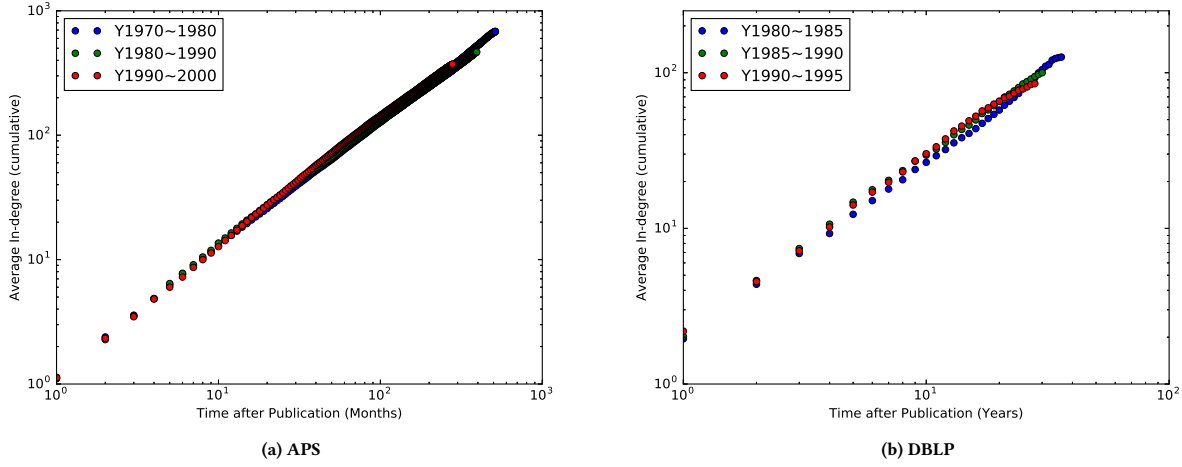

(a) APS

(b) DBLP

Figure 4: The average growth curve of number of citations of papers published in different time periods in (a) APS and (b) DBLP. Papers in the two datasets are grouped according to their publish year (papers with citation fewer than 20 are excluded).

## 4 MODEL

In this section, we propose our model as an extension of the Bianconi-Barabási model and the relevance model (see Section 2.4). Based on the above described empirical observations on real-world data, we specify the following three assumptions of our model at an abstract level.

- Preferential attachment with decay of relevance;
- Exponential growth of the network size;
- Power-law degree growth of nodes with regard to their ages.

### 4.1 Network Generation

We now demonstrate the process of network generation using our model. The generative model takes three parameters:

- $\alpha$: the exponential growth rate of the network size;
- $m$: the number of links a node creates at its arrival;
- $\rho(\eta)$: the fitness distribution.

*4.1.1 Preferential Attachment with Decay of Relevance.* Similar to the models in Section 2, our model starts with one node with a self loop (thus having indegree $k_1 = 1$) at the initial event time $s = 1$. The first node is assigned a fitness value $\eta$ sampled from the fitness distribution $\rho(\eta)$.

At each later step, a new node joins the network with a self loop and is also assigned a fitness value $\eta$ sampled from $\rho(\eta)$. The new node creates a constant number of $m$ links to existing nodes. The probability $\Pi_i$ that a new link connects to an existing node $i$, is proportional to the product of its degree $k_i$, its fitness $\eta_i$ and a decay function value $f_R(\tau_i)$ that depends on its age $\tau_i$.

Since the new node creates $m$ links, for an existing node $i$ the growth of its degree $\frac{\partial k_i}{\partial s}$ with respect to the event time $s$ is thus $m$ times its preferential attachment probability $\Pi_i$:

$$\frac{\partial k_i}{\partial s} = m\Pi_i = m \frac{k_i \eta_i f_R(\tau_i)}{\sum_j k_j \eta_j f_R(\tau_j)} \quad (8)$$

*4.1.2 Exponential Growth of the Network Size.* We follow observations about network growth made in Section 3.3 and model network growth to be exponential over physical time $t$. We realize this by defining event time $s$ to proceed by one unit every time a new node joins in. This leads to the following exponential relationship between $t$ and $s$:

$$s = e^{\alpha t} \quad (9)$$

### 4.2 Analytical Solution of the Decay Function $f_R(\tau)$

The last assumption in our model, the power-law growth of node degrees:

$$k_i(\tau_i, \eta_i) = \tau_i^{\beta(\eta_i)} \quad (10)$$

does not appear in the generative process. Instead, it is used to derive the decay function $f_R(\tau)$. Later we evaluate the degree growth in Section 5.2.

We first calculate the denominator $\sum_j k_j \eta_j f_R(\tau_j)$ in Equation 8. When the age of the network $T$ is large enough, applying the continuum approximation we get:

$$\sum_j k_j \eta_j f_R(t_j) = \int d\eta \rho(\eta) \eta \int_1^T d\tau k(\tau, \eta) f_R(\tau) \cdot \alpha e^{\alpha(T-\tau)} \quad (11)$$

Intuitively, the tern $e^{\alpha(T-\tau)}$ says there are more nodes with smaller ages in the system. Now let us assume that the decay function has a form of $f_R(\tau) = \tau^{-1}$. Plugging in Equation 10, for a certain fitness $\eta$ we have:

$$\int_1^T d\tau k(\tau,\eta)f_R(\tau) \cdot \alpha e^{\alpha(T-\tau)} = \alpha e^{\alpha T} \int_1^T d\tau \frac{\tau^{\beta(\eta)-1}}{e^{\alpha\tau}} \quad (12)$$

$$= \alpha e^{\alpha T} \cdot \frac{\Gamma(\beta(\eta),\alpha)}{\alpha^{\beta(\eta)}} \quad (13)$$

where $\Gamma(a,b) = \int_b^{+\infty} x^{a-1}e^{-x}\,dx$ is the upper incomplete gamma function. Thus, the denominator of Equation 8 when $T \to +\infty$ is proportional to the network size $e^{\alpha T}$:

$$\lim_{T \to +\infty} \sum_j k_j \eta_j f_R(\tau_j) = C e^{\alpha T}, \quad (14)$$

with the constant

$$C = \alpha \int d\eta \rho(\eta)\eta \cdot \frac{\Gamma(\beta(\eta),\alpha)}{\alpha^{\beta(\eta)}}. \quad (15)$$

Recalling Equation 8 we get

$$\frac{\partial k_i}{\partial s} = m\Pi_i = m\frac{k_i\eta_i f_R(\tau_i)}{\sum_j k_j\eta_j f_R(\tau_j)} = m\frac{k_i\eta_i}{Ce^{\alpha T}\tau} = \frac{\partial k_i}{\partial t} \cdot \frac{dt}{ds} \quad (16)$$

We therefore have

$$\frac{\partial k_i}{\partial t} = \frac{\alpha m \eta_i k_i}{C\tau}, \quad (17)$$

which has an analytical solution that recovers Equation 10:

$$k_i(\tau_i,\eta_i) = \tau_i^{\beta(\eta_i)} \quad (18)$$

given $\beta(\eta_i) \sim \eta_i$ (as in the Bianconi-Barabási model):

$$\beta(\eta_i) = \frac{\alpha m \eta_i}{C}, \quad (19)$$

and the initial value of $k_i$ being 1:

$$k_i(1,\eta_i) = 1. \quad (20)$$

To summarise, the decay function $f_R(\tau) = \tau^{-1}$ is a function that let the model be self consistent.

*4.2.1 Other Forms of Decay Functions.* Other forms of decay functions that differ from our solution such as the exponential decay have been suggested [12]. We now explain why an exponential decay function having the form $f_{R(\exp)} = e^{-\gamma\tau}$ will not let Equations 8, 9 and 10 be self consistent.

Rewriting the denominator of Equation 8 with $f_{R(\exp)}$ we find that when $T \to +\infty$ the denominator is still proportional to the network size $e^{\alpha T}$:

$$\lim_{T \to +\infty} \sum_j k_j\eta_j f_R(\tau_j) = C_{\exp} \cdot e^{\alpha T}, \quad (21)$$

with the constant

$$C_{\exp} = \alpha \int d\eta \rho(\eta)\eta \cdot \frac{\Gamma(\beta(\eta)+1,\alpha+\gamma)}{(\alpha+\gamma)^{\beta(\eta)+1}}. \quad (22)$$

Thus, the counterpart of Equation 17:

$$\frac{\partial k_i}{\partial t} = \frac{\alpha m \eta_i k_i \cdot e^{-\gamma\tau}}{C_{\exp}} \quad (23)$$

leads to:

$$k_i = \exp(-\frac{\alpha m \eta_i e^{-\gamma\tau}}{\gamma C_{\exp}} + c) \quad (24)$$

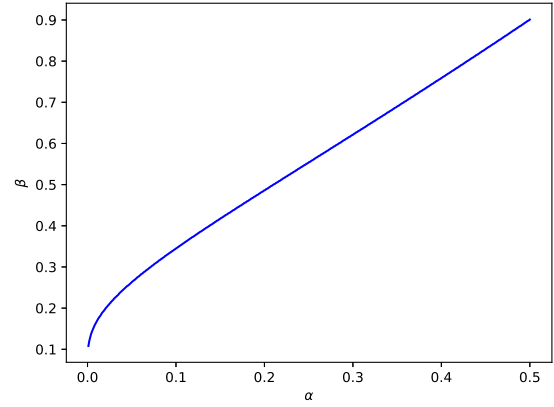

Figure 5: The relationship between $\alpha$ and the power-law exponent $\beta$ of the degree growth when we take $m = 1$.

which is a double exponential function and does not recover Equation 10. Whether the exponential relevance decay recovers other forms of degree growth functions is however an open question.

### 4.3 Theoretical Degree Distribution

We now give the theoretical stationary degree distribution of our model. When we have a constant fitness $\eta$ for all nodes, the theoretical degree distribution when $T \to +\infty$ will converge to:

$$p(k) = \frac{\alpha}{\beta} \cdot k^{-1} \cdot e^{-\alpha k^{1/\beta}} \cdot k^{1/\beta} \quad (25)$$

which is a product of a power function and a stretched exponential function.

When we have a distribution of fitness the degree distribution will be a superposition of Equation 25 with different $\beta$ values corresponding to different $\eta$ values:

$$p(k) = \alpha k^{-1} \int d\eta \rho(\eta) \cdot \beta(\eta)^{-1} e^{-\alpha k^{1/\beta(\eta)}} \cdot k^{1/\beta(\eta)} \quad (26)$$

Particularly if the fitness distribution follows a Zipf's law i.e., $\rho(\eta) \sim \eta^{-1}$, Equation 26 reduces to:

$$p(k) \sim \frac{e^{-\alpha k}}{k \cdot \ln k} \quad (27)$$

Note that for our model we do not get a perfect power-law distribution as in the Barabási-Albert model. However one should keep in mind that in empirical data, it is often hard if not possible to discern whether the observed data is closer to a power-law or a similar distribution with a heavy tail [6], thus we cannot presuppose every distribution with a heavy tail to be power-law.

### 4.4 Slowing-down Growth of the Network Size

Now we discuss the situation where the exponential growth of the network size slows down. In our model, such slowdown can be realised by varying the parameter $\alpha$. Suppose we have a constant fitness $\eta$, the relationship between $\alpha$ and the power-law exponent

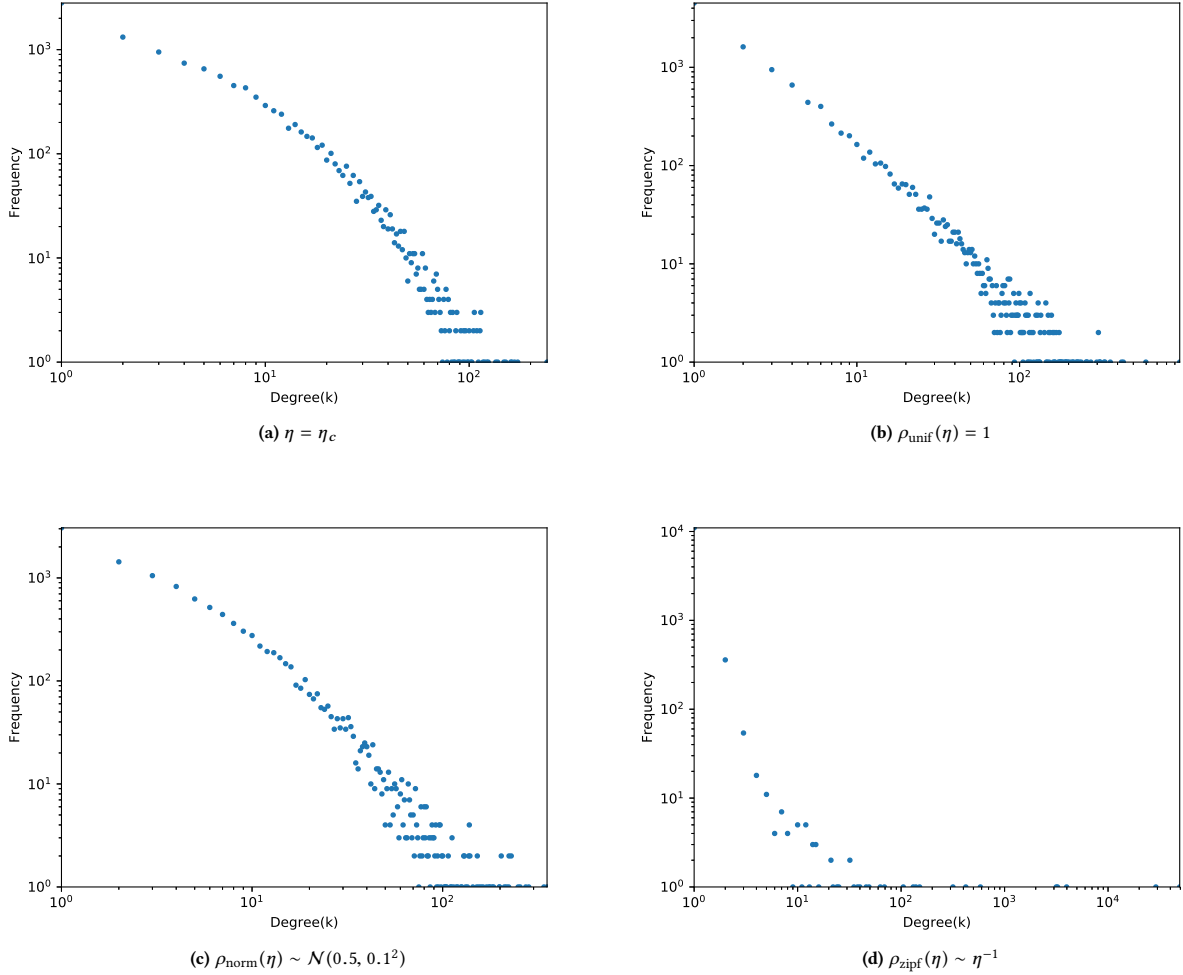

Figure 6: The degree distributions of the synthetic networks generated with different fitness distributions.

of degree growth $\beta$ can be obtained from Equation 15 and 19:

$$\beta = \frac{\alpha m \eta}{C} = \frac{m\alpha^\beta}{\Gamma(\beta, \alpha)} = \frac{m}{E_{1-\beta}(\alpha)}, \quad (28)$$

where $E_{1-\beta}(\alpha)$ is the generalised exponential integral function [14]. Numerically we show the relationship between $\alpha$ and $\beta$ in Figure 5. As we can see, $\beta$ is monotonically decreasing as $\alpha$ decreases. This indicates the slowdown of the growth of the network size causes the slowdown of the degree growth of individual nodes. In the extreme case when $\alpha$ falls down to zero, the network size stops growthing completely, and there is no degree growth of individual nodes as well.

## 5 EVALUATION

In this section, we use our model to generate synthetic networks with different parameters in order to evaluate its plausibility. At the macro level, we evaluate the degree distributions of the generated networks. At the micro level, we evaluate the assumption we have made but not directly reflected in the generative model: the power-law growth of node degrees.

### 5.1 Degree Distribution

Now we generate synthetic networks with our model using different fitness distributions $\rho(\eta)$ and observe the resulting degree distributions of the network. The four fitness distributions we look at are:

- constant fitness
$$\eta = \eta_c \quad (29)$$

- uniform distribution
$$\rho_{\text{unif}}(\eta) = 1, \qquad \eta \in (0, 1) \quad (30)$$

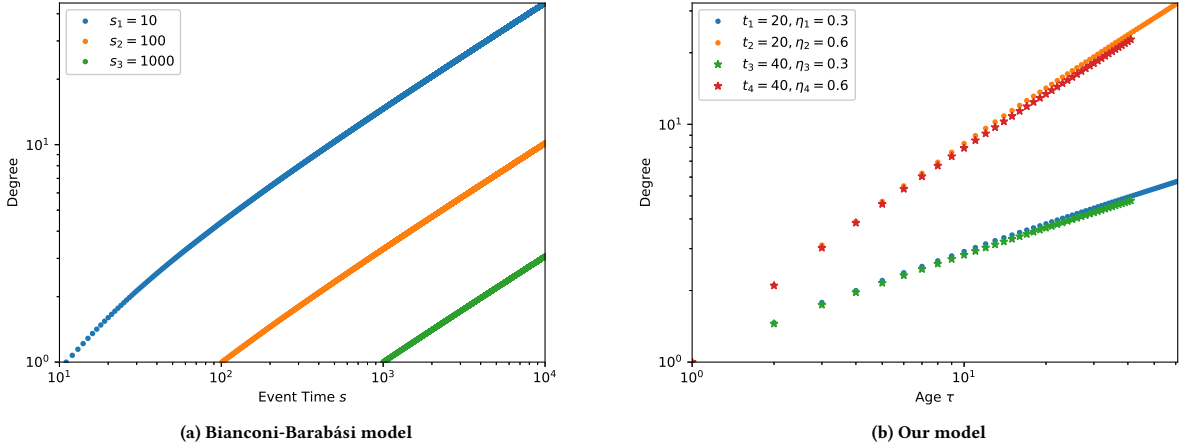

Figure 7: The degree growth curves of nodes with different parameters in the synthetic network generated by (a) the Bianconi-Barabási model, and (b) our model. Note the difference in the X-axes: In (a) it represents the event time $s$, while in (b) it represents the node age $\tau$ in physical time.

- normal distribution
$$\rho_{\text{norm}}(\eta) \sim \mathcal{N}(0.5, 0.1^2), \qquad \eta \in (0,1) \quad (31)$$
- Zipf's law distribution
$$\rho_{\text{zipf}}(\eta) \sim \eta^{-1}, \qquad \eta \in (0,1) \quad (32)$$

Other parameters of our model are set to $\alpha = 0.1$ and $m = 8$. We plot the degree distributions of the generated networks at $T = 70$ in Figure 6.

The curves in all four subfigures decay slower than exponentially as the degree $k$ grows. In Figure 6a an obvious cut-off can be observed. In Figures 6b and 6c the degree distribution curves well reflect what we have observed empirically in Figure 1. In Figure 6d we observe a strong imbalance in the degree distribution: Few nodes are so well connected ($k > 10^3$) while the majority of nodes have degrees less than 10. This is due to the fact that in the case where we have a Zipf's law distribution of fitness, the "rich-get-richer" effect has become so dominant that it turns into "winner-takes-all", akin to the Bose-Einstein condensation [2] in the Bianconi-Barabási model.

### 5.2 Power-law Degree Growth

The degree growth of a node is power-law in both our model and the Bianconi-Barabási model [3], but with different bases. The latter predicts Equation 7: $k_i \sim (s/s_i)^{\beta(\eta_i)}$, which means that the degree growth of a node $i$ does not only depend on its age, but also the time it joins the network $s_i$ itself. Therefore, nodes who join the network earlier have faster degree growth rates than those who join later, as known as the first-mover advantage. This is due to the fact that the Bianconi-Barabási model does not consider the decay of node relevance over time. In our model however, the degree growth of a node $i$ solely depends on its fitness $\eta_i$ and its age $\tau_i$, but not the time it joins the network, i.e., $k_i \sim \tau_i^{\beta(\eta_i)}$, as in Equation 10.

To illustrate the difference numerically, we generate synthetic networks[4] using both models, and plot the degree growth curves of different nodes in Figure 7. We choose a uniform fitness distribution as in Equation 30.

Figure 7a shows exactly what the Bianconi-Barabási model predicts. We see the degree growth curves of three nodes which joined at timestep $s_1 = 10$, $s_2 = 100$ and $s_3 = 1000$ respectively, and have the same fitness $\eta = 0.5$. The three parallel straight lines suggest that their degree growths follow power-law with the same exponent. However, node 2 needs ten times longer time to get the same degree of node 1, and node 3 needs 100 times longer, matching Equation 7 in the Bianconi-Barabási model.

We see in Figure 7b that in our model, nodes with the same fitness ($\eta_1 = \eta_3 = 0.3$, while $\eta_2 = \eta_4 = 0.6$) have comparable exponents of the power-law degree growth, regardless of the time when they joined the network ($t_1 = t_2 = 20$, while $t_3 = t_4 = 40$). Besides, nodes with higher fitness have faster degree growth, and the exponents are proportional to the fitness values. These observations match Equation 10 in our model.

Comparing with the empirical observations in real datasets (Figures 3 and 4), we find that our model has a more realistic power-law degree growth function.

## 6 CONCLUSION

We have proposed a new preferential attachment–based network growth model. Our model connects the macro and micro levels of phenomena in evolving networks: the growth of network sizes, the growth of node degrees and the decay of node relevance, leading to deeper thoughts about the co-existence of the two mechanisms in networks: decay and growth.

---
[4]To avoid the nondeterminism in micro actions caused by discreteness, here we allow degrees to have continuous values.

We have been focusing on citation networks where link creation is only allowed at the arrival of new nodes. The problem posed in our model can be generalised to other networks on the Web which allow the creation or even the removal of links between existing nodes, and lead to more interesting yet more challenging studies.

## ACKNOWLEDGMENTS

The authors would like to thank the APS for providing the dataset. The research leading to these results has received funding from the European Community's Horizon 2020 - Research and Innovation Framework Programme under grant agreement n$^o$ 770469, CUTLER.